\documentclass[prl,a4paper,twocolumn,showpacs]{revtex4}
\usepackage{graphicx}
\newcommand{\be}{\begin{equation}}
\newcommand{\ee}{\end{equation}}
\newcommand{\bea}{\begin{eqnarray}}
\newcommand{\eea}{\end{eqnarray}}
\newcommand{\ba}{\begin{array}}
\newcommand{\ea}{\end{array}}
\newcommand{\ben}{\begin{enumerate}}
\newcommand{\een}{\end{enumerate}}
\newcommand{\bei}{\begin{itemize}}
\newcommand{\eei}{\end{itemize}}

\newcommand{\e}{\epsilon}

\newcommand{\om}{\omega}

\newcommand{\s}{\sigma}
\newcommand{\bd}{b^\dagger}
\newcommand{\cd}{c^\dagger}
\newcommand{\vd}{v^\dagger}

\newcommand{\dek}{\Delta\epsilon_k}
\newcommand{\dekt}{\Delta\epsilon_{k'}}

\newcommand{\la}{\langle}
\newcommand{\ra}{\rangle}

\newcommand{\lm}{\lambda}
\newcommand{\lmq}{{q\lambda}}
\newcommand{\D}{\Delta}
\newcommand{\ua}{\uparrow}
\newcommand{\da}{\downarrow}

\begin{document}
\title{Manipulation of collective spin correlations in semiconductors with polarization squeezed vacuum}
\author{E. Ginossar, Y. Levinson and S. Levit}
\affiliation{Department of Condensed Matter Physics, The Weizmann Institute of Science, Rehovot 76100, Israel}
\email{eran.ginossar@weizmann.ac.il}
\date{\today}
\begin{abstract}
We calculate the transfer rate of correlations from polarization
entangled photons to the collective spin of a many-electron state in a two-band system.
It is shown that when a semiconductor absorbs pairs of photons from a two-mode squeezed vacuum,
certain fourth order electron-photon processes correlate the spins of the excited electron pairs of different quasi-momenta.
Different distributions of the quantum Stokes vector of the light lead to either enhancement or reduction of the
collective spin correlations, depending on the symmetry of the distribution.
We find that as the squeezing of the light becomes non-classical, the spin correlations exhibit a crossover from
being positive with a $\sim N^2$ ($N$ is average photon number) scaling, to being negative with $\sim N$ scaling,
even when $N$ is not small. Negative spin correlations mean a preponderance of spin singlets in the optically generated state.
We discuss the possibility to measure the collective spin correlations in a
combined measurement of the Faraday rotation fluctuation spectrum and excitation density in a steady-state configuration.
\end{abstract}
\maketitle

Optical excitation of a semiconductor with circularly polarized light generates an average
collective spin polarization in the conduction band \cite{optical-orientation,dymnikov,Levinson}.
This allows to investigate spin relaxation mechanisms in semiconductors using techniques such as
 time-resolved Faraday rotation, time-resolved photoluminescence, and femto-second pulses \cite{fabian-rmp,pfaltz}.
It is also possible to monitor the position of the
electron spins, their phase and amplitude as well as coherently control them \cite{awschalom, oestreich}.
Recently  the fluctuations of electronic collective spin of a bulk GaAs sample
 were measured \cite{Oestreich-2}. Theoretically, spin states are interesting because of their
 relation to squeezing \cite{kitagawa,kuzmich,Palma-Knight,Agarwal-Puri} and to entanglement \cite{lukin}.
It has been suggested that collective atomic spins can store entanglement, optically transferred
from correlated photons to the atomic cloud \cite{hammerer,serafini}.

In this work we study how the spin correlations in a semiconductor system can be optically
generated in a controlled way.
We consider a pump beam of correlated photons which is absorbed in a semiconductor,
exciting a density of electron-hole pairs \cite{paper}.
Absorbed photons   are found to either enhance or reduce  the spin-spin correlations depending
on the correlations of optical modes with different wavelengths and either the same or different polarizations.
In particular we find that it is necessary to use squeezed  light in its non classical regime in order to excite net anti-correlated spins.
 This is due to a competition between two fourth-order processes inducing positive and
negative correlations, the latter becoming dominant only for
non-classical light.  We note  that  net spin anticorrelations mean a preponderance of singlet spin components.

Polarization properties of photons are described by the quantum Stokes parameters \cite{collet-book,korolkova}
which in the circular polarization basis ($\epsilon_{\pm}$)are written as, Ref.\cite{jackson-book},
\be \label{stokes-vec-def} \hat{p}_i=\left(\ba{cc}\epsilon_+\cdot\bf{E}^{\dagger}\,& \epsilon_-\cdot\bf{E}^{\dagger}\ea\right) \s_i
\left(\ba{c}\epsilon^*_+\cdot\bf{E}\\  \epsilon^*_-\cdot\bf{E}\ea\right)
     =\sum_{q,q'} \vec{b_{q}}^{\dagger}\s_i \vec{b_{q'}} \ee
where $\textbf{E}$ is the  electric field $\textbf{E}=\sum_{q\lm} \hat{\epsilon}_\lm b_{q\lm}$ at position $r=0$
, $\vec{\bd_{q}}=(\bd_{q+},\bd_{q-})$ and $\s_{i=0..3}$ denote the unit and Pauli spin matrices.
The operators $b_\lmq$ are the photon annihilation operators  with wave number $q$ and  polarization $\lm$.
We consider a collinear pump  beam  with a range of frequencies $\om_0\pm B/2$ above the electron-hole gap.
We assume that the average Stokes parameters in a given bandwidth $B$ are $\la \hat{p}_{0}\ra=\frac{2\pi c}{LB}\sum_{q\lm}N_{q\lm}$,
$\la \hat{p}_{1}\ra =\la \hat{p}_{2}\ra=\la \hat{p}_{3}\ra =0$, where $N_{q\lambda}$ is is the average photon occupation per mode and
$L$ is the quantization length. This describes an  unpolarized light with  the Stokes vector fluctuating around the origin of the Poincar\'{e} sphere.
The fluctuations are described by the covariance matrix $p_{ij}=1/2\la \hat{p}_i \hat{p}_j+\hat{p}_j\hat{p}_i \ra$, which for a Gaussian type field depends
on the normal $\la \bd_{q\lm} b_{q'\lm'}\ra$ as well as anomalous $\la b_{q\lm} b_{q'\lm'}\ra$ correlations, the latter
constituting the main characteristics of squeezed vacuum \cite{collet-loudon}. Beams with such properties
 are generated  using the  parametric down-conversion, \cite{shih-1,grice}.
In addition to normal correlations  $\la \bd_{q\lm}b_{q'\lm'} \ra=N_q\delta_{\lm \lm'}\delta_{q q'}$ they possess two generic anomalous correlations:
 same polarization squeezing $\la b_{q\pm}b_{q'\pm} \ra=M^{(1)}_{q\pm}\delta_{q+q',2q_0}\delta_{\om_q+\om_{q'},2\om_0}$
and opposite polarization squeezing $\la b_{q\pm}b_{q'\mp} \ra=M^{(2)}_{q\pm}\delta_{q+q',2q_0}\delta_{\om_q+\om_{q'},2\om_0}$,
where $M^{(1,2)}_{q\pm}$ are complex functions.
Since for the squeezed vacuum $\la \hat{p}_0 \hat{p}_{1,2,3}\ra=0$, the fluctuations of $\hat{p}_{1,2,3}$ can be described separately from the variance of $\hat{p}_0$.

\begin{figure}[h]
\centering
\includegraphics[scale=0.6]{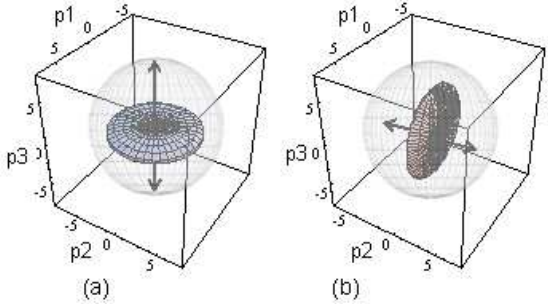}
\caption{Two generic fluctuation patterns of the Stokes vector ($p_1,p_2,p_3$). In the circular basis ($\pm$),
they are described by covariance ellipsoids for (a) opposite-polarization squeezing (for average
parameters $N=1,M_{\pm}^{(1)}=0,M_{\pm}^{(2)}=1.314$) and (b) same-polarization squeezing
($N=1,M^{(1)}_{\pm}=1.314,M^{(2)}_{\pm}=0$). The value of $M$ was chosen to be slightly below the maximal
squeezing $\sqrt{N(N+1)}$.}
\label{ellipsoids-figure}
\end{figure}

It is instructive to draw  the covariance ellipsoids for the tensor $p_{ij}$.  Such ellipsoids are shown in
Fig. \ref{ellipsoids-figure} for the two cases of the same-polarization and opposite-polarization squeezing
based on averaged occupation and squeezing functions $\bar{N}_q=N$ and $\bar{M}^{(1,2)}_{q\pm}=M^{(1,2)}_{\pm}$.
On the axes are plotted $p_i$,  possible values of the averages $\la \hat{p}_i\ra$.
The variance $p_{33}$ given by $2[N(N+1)+ |M^{(1)}|^2 \pm |M^{(2)}|^2]$ for
$|M^{(1,2)}_{+}|=|M^{(1,2)}_{-}|$ indicating that correlations of the type $M^{(1)}$ ($M^{(2)}$) enhance
(reduce) the variance $p_{33}$, a fact which is important for spin-spin correlations.

The free part of the Hamiltonian of the semiconductor is modelled as a two-band system
\be\label{H_0}
H_0=\sum_{q,\lm}\omega_\lmq\bd_\lmq b_\lmq +\sum_{k\s}\epsilon_{k\s}^c\cd_{k\s}c_{k\s}+\sum_{k\s}\epsilon_{k\s}^v\vd_{k\s}v_{k\s}. \ee
The operators $c_{k\s}$ and $v_{k\s}$ denote annihilation operators of the free electrons in the conduction and valence
 bands, with quasi-momentum $k$ and spin $\s$.
The interaction Hamiltonian of the electrons and the photons in the dipole approximation is given by \cite{Haug-Koch-Book,Khitrova-Review}

\bea \nonumber \label{V_Int}
&& V=\sum_{\s,\s',\lm}\sum_{p,q}\left[A_{pq}^{\s \s' \lm}\cd_{p+q\s}v_{p\s'}b_\lmq + h.c. \right]
\eea
where $A_{pq}^{\s\s'\lm}$ are the interaction matrix elements for
 dipole transitions from heavy-hole band to conduction band in GaAs near the $\Gamma$ point \cite{optical-orientation}.

In the present work we will not consider the  electron-electron and electron-phonon interactions.
 The quantum optical effects on the rate of generation which
we will describe in the context of the simple two band model should qualitatively hold also in the presence of the interactions. We will make more
detailed comments supporting this assumption later on.

For the optical  beam described above we wish to calculate the rate at which it is generating
collective spin correlations in a semiconductor. The average total spin $\la \textbf{S} \ra$ of the photo-excited conduction
electrons is zero  since the squeezed vacuum radiation is unpolarized. Consider the spin-noise two time average
\be \label{spin-flucs} \la \textbf{S}(t)\cdot\textbf{S}(t')\ra=\sum_{k,k'}\la \mathbf{\mathcal{S}}_k(t)\cdot\mathbf{\mathcal{S}}_{k'}(t')\ra \ee
where $\mathcal{S}_k=\sum_{\s\s'}\cd_{k\s}\vec{\s}_{\s\s'}c_{k\s}$. It describes the fluctuations of the total spin $\textbf{S}$ and  consists of two parts - the
sum of fluctuations of individual spins $\sum_k \la \mathbf{\mathcal{S}}_k(t)\cdot\mathbf{\mathcal{S}}_k(t') \ra$ and the collective pairwise correlations, i.e. terms with
$k\ne k'$. Their {\em variance} is given by $\la \bf{S}^2\ra - \sum_k \la \mathbf{\mathcal{S}}_k^2 \ra$.

To lowest order  in the dipole interaction the contribution to the collective correlations  comes from the optically induced transition of two electrons
into states $k,\s$ and $k',\s'$. Physically, we expect that the two excited spins will be positively (negatively) correlated if the two absorbed photons are correlated and
have the same (opposite) polarizations. Indeed the main result, Eq. (\ref{c-2}) below,  of the calculation reflects the competition between the strengths of the
correlations existing in the photon field: an auto-correlation within each mode, and a cross-correlation between different modes
due to the squeezing.

The second order contribution for the total spin fluctuations is given by
\small
\bea \label{s2-general-expr}\nonumber
&& \la S^2\ra^{(2)}= \\ \nonumber
&&= \sum_{\{1..4\}}
 A_{p_1,q_1}^{\s_1,\s'_1,\lm_1*} A_{p_2,q_2}^{\s_2,\s'_2,\lm_2*}C_{p_1,q_1}^{p_2,q_2*}
 A_{p_3,q_3}^{\s_3,\s'_3,\lm_3} A_{p_4,q_4}^{\s_4,\s'_4,\lm_4}C_{p_3,q_3}^{p_4,q_4} \times \\
&&\times \la \bd_{q_1\lm_1}\bd_{q_2\lm_2}b_{q_4\lm_4} b_{q_3\lm_3}\ra_{rad}
\la 1,2| S^2 |3,4\ra_{eq}
\eea\normalsize
where  $\la\bd_{q_1\lm_1}\bd_{q_2\lm_2}b_{q_4\lm_4} b_{q_3\lm_3}\ra_{rad}$ is a property of the external field,
and we define
\bea
&& C_{p_1,q_1}^{p_2,q_2}=\frac{e^{i(\D\e_{p_1q_1}+\D\e_{p_2q_2}-\om_{q_1}-\om_{q_2})t+2\eta t}}{\D\e_{p_1q_1}+\D\e_{p_2q_2}-\om_{q_1}-\om_{q_2}-2i\eta} \times \\ \nonumber
&& \times \left[\frac{1}{\D\e_{p_1q_1}-\om_{q_1}-i\gamma_{p_1}}+\frac{1}{\D\e_{p_2q_2}-\om_{q_2}-i\gamma_{p_2}} \right] \eea
which is the second order amplitude, where $e^{\eta t}$ the adiabatic switching on factor, and  $\gamma_p$ is
the lifetime  of the conduction electron state \cite{heitler,kubo-inter}.
In expression (\ref{s2-general-expr}) $\la 1,2| S^2 |3,4\ra$ is the fermionic average
\small\bea
&& \la 1,2| S^2 |3,4\ra=\sum_i \sum_{k,s_1,s_2}\sum_{k',s'_1,s'_2}\s^{(i)}_{s_1,s_2}\s^{(i)}_{s'_1,s'_2}\times \\ \nonumber
&& \times \la c_{p_1+q_1\s_1}c_{p_2+q_2\s_2}\cd_{ks_1}c_{ks_2}\cd_{k's_1'}c_{k's_2'}
\cd_{p_3+q_3\s_3}\cd_{p_4+q_4\s_4} \ra_{eq} \times \\ \nonumber
&& \times\la \vd_{p_1\s_1'}\vd_{p_2\s_2'}v_{p_3\s_3'}v_{p_4\s_4'} \ra_{eq} \eea \normalsize
where $\la\ra_{eq}$ is assumed to be equilibrium at $T=0$.

\begin{figure}[h]
\centering
\includegraphics[scale=0.6]{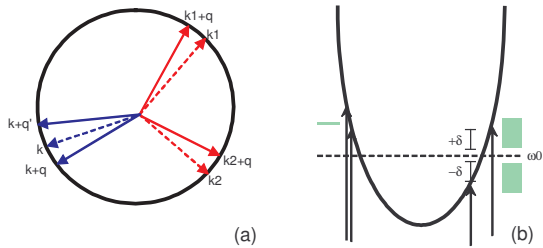}
\caption{Correlation processes in the $k$-space. (a) The excitation of a singlet pair of momentum $k$
(blue dashed arrow) from the valence band into the conduction band (blue solid arrow) and excitation
of two valence band electrons of different momenta (red). (b) Pair excitations associated with normal (left) photonic contractions
 (left - narrow green strip) and excitations associated with anomalous correlations (right - wide green strips).}
\label{processes-figure}
\end{figure}

A Wick decomposition of expression (\ref{s2-general-expr}) contains contractions which contribute  to the
independent fluctuations $\sum_k\la \mathcal{S}_k^2\ra $ as well as contractions which contribute to the collective
spin-spin correlations $\la \mathcal{S}_k\cdot \mathcal{S}_{k'}\ra$ with $k\neq k'$.
The latter can be further divided \footnote{These two processes are different contractions in (\ref{s2-general-expr}).} into two processes,
Fig. \ref{processes-figure}a, in which (i) a singlet ($k\uparrow,k\downarrow$) in the valence band is broken
into two different momenta in the conduction band ($k+q,k+q'$) and (ii) two electrons with
different momenta ($k_1,k_2$) are excited into the conduction band with momenta ($k_1+q,k_2+q'$).
Process (i) has  considerably smaller rate with respect to (ii) because most of the phase space of final
states cannot be reached with the typically small photon momentum.
The ratio can be approximately estimated to be $(\frac{B}{ck})^2$  where $B$ is the optical bandwidth, and $k$ is
the typical electron wave number. Therefore in the following we neglect the contribution of process (i).

For the generation rate of correlations due to process (ii),
we use the corresponding contractions in (\ref{s2-general-expr}), differentiating
with respect to time and taking the  limit $\eta\rightarrow 0$. This results in an energy conservation constraint
for the entire process of exciting two electron-hole pairs.
This process again has two parts (Fig. \ref{processes-figure}b): one coming from normal contractions
$\la\bd b\ra^2$   and another from anomalous contractions $|\la b b \ra|^2$ due to squeezing.
For normal contractions any two spin components  $\mathcal{S}_k,\mathcal{S}_{k'}$ become correlated due to the
absorption of two photons from the same mode $q$, obeying the energy conservation $\om_q=\frac{1}{2}(\dek+\dekt)$.
In contrast, for anomalous contractions only spin components which have symmetric energies $\dek+\dekt=2\om_0$ become correlated.
These are drawn out of a continuum of such pairs obeying $\om_q+\om_{q'}=2\om_0$.
Therefore the two processes are distributed very differently
in phase space, although they have the same total phase space.
These processes give the largest contribution to the generation rate of correlations
$\la \bf{S}^2\ra - \sum_k \la \mathbf{\mathcal{S}}_k^2 \ra$, and in the limit of $q\ll k$ are given by (per unit volume)
\bea \label{c-1}
&& C_s = C_{s0}\sum_{q}\int d\dek \, \rho(\dek)\rho(2\om_0-\dek)\times \\ \nonumber
&& \times \left(N^2_q+|M^{(1)}_{q}|^2-|M^{(2)}_{q}|^2\right)\frac{\gamma^2}{\left[(\dek-\om_{q})^2+(\frac{\gamma}{2})^2\right]^2} \\ \nonumber
\eea
where $C_{s0}=\frac{32\pi^2|d|^4}{3\hbar}$ with $d$ the dipole matrix element, and we assume $|M^{(1,2)}_{q\pm}|=|M^{(1,2)}_{q}|$ i.e. that the ($\pm$) squeezing
correlations differ only by phase. The numerical factors in $C_{s0}$ is due to angular integrals and a symmetry factor
of the contraction.
\begin{figure}[h!]
\centering
\includegraphics[scale=0.4]{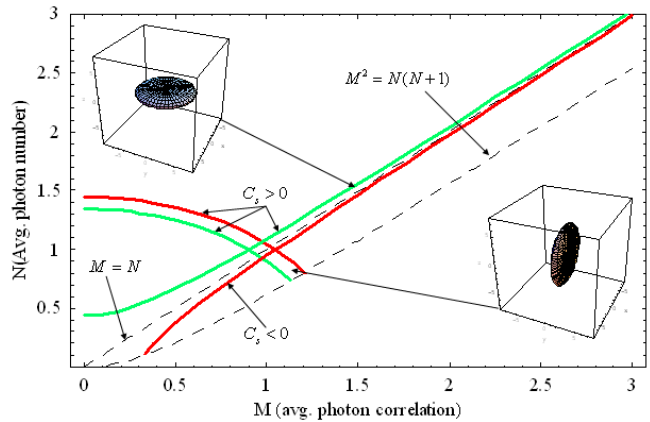}
\nonumber
\caption{Curves of equal collective spin correlation in the parameters of average occupation ($N$) and squeezing ($M$).
Cases drawn for enhanced $M^{(2)}=0$ (circle quadrants) and reduced $M^{(1)}=0$ (hyperbola quadrants) spin correlations. The colors
signify different values of $C_s$. Inserts: Patterns of fluctuations of the Stokes vector leading to the enhanced  and reduced
 spin correlations.}
\label{resultfig}
\end{figure}
We see from expression (\ref{c-1}) that away from the edges of the bandwidth the correlation per unit bandwidth is
simply proportional to $N^2+|M^{(1)}|^2-|M^{(2)}|^2$ (assuming $N_q \simeq N$ and $M^{(1,2)}_{q}\simeq M^{(1,2)}$ across most of the bandwidth).
Since the  electronic
density of state in the excitation bandwidth can be taken as constant, integrating (\ref{c-1}) with respect to the electronic energy $\dek$
gives
\be \label{c-2} C_s = C_{s0} \rho(\om_0)^2\frac{4\pi}{\gamma}\sum_{q}
\left(N^2_q+|M^{(1)}_q|^2-|M^{(2)}_q|^2 \right).\ee

This result shows that increased (decreased) fluctuations of the Stokes parameter $p_3$ cf. Fig \ref{ellipsoids-figure}, lead to increased (decreased) spin correlations.
Positive spin correlations induced by absorbing photons from the same mode are enhanced by absorbing squeezed photons with the same polarization
modes and reduced by the squeezing of the opposite polarizations, cf. Fig. \ref{resultfig}.
The total spin correlations can become negative when the squeezing
is {\em non-classical}, i.e. $M^{(2)}>N$. The maximal negative spin correlations will be reached for a pure squeezed state
with $M^{(2)}=\sqrt{N(N+1)}$, Ref. \cite{Gardiner86}, in which case $C_s \sim -N$.
Remarkably this is a rare case in which a quantum optical effect is not confined to small photon occupations
$N \ll 1$.

It is instructive to consider the equal spin correlation curves, Fig. \ref{resultfig} which indicate  that the geometry of negative
correlation curves is separated from the other cases: the hyperbolic curves are confined to either the classical side ($C_s>0$) or the
quantum side ($C_s<0$) of the diagram, with the $C_s=0$ being a separatrix between the two regimes. The possibility to completely
eliminate  the  inevitable spin correlations induced by unsqueezed light (the first term in Eq. (\ref{c-2})) may be useful for observing other sources
of spin correlations  such as contributions from nuclear spins.

It is useful to define a reduced density matrix $\rho^{(k,k')}_{\alpha \beta}$ for a pair of spins $\mathcal{S}_k,\mathcal{S}_{k'}$,
where $\alpha,\beta$ run over the singlet ($|0,0\ra$) and three triplet basis states ($|1,0\ra,|1,\pm 1\ra$). In this basis it can
be easily shown that for negative spin correlations $C_s<0$ the diagonal elements $\rho_{\alpha}\equiv \rho_{\alpha,\alpha}$ obey
\be \sum_{k,k'}\rho^{(k,k')}_{0,0}>\frac{1}{3}\sum_{k,k'}(\rho^{(k,k')}_{1,-1}+\rho^{(k,k')}_{1,0}+\rho^{(k,k')}_{1,1})\ee
which means that for the electronic state generated by non-classical light, there is a preponderance of the singlet component in the pairwise spin correlations.

An enhancement or reduction of spin-spin correlations should be measurable
from the difference of $\la \textbf{S}^2\ra$ with squeezed and unsqueezed light, as can be seen from Eq. (\ref{c-2}).
It is also in principle  possible to observe the spin correlations by measuring  $\la \textbf{S}^2\ra $ and
$\sum_k \la \mathbf{\mathcal{S}}_k^2 \ra$. The total spin fluctuations $\la \textbf{S}^2 \ra$ can be
estimated from the
variance of the magnetic moment of the sample for example in a Faraday rotation setup similar to the one used to measure thermal
spin fluctuations \cite{Oestreich-2}. For the diagonal part of the fluctuations we can use the identity
\be \sum_k \la S_k^2 \ra = 3\sum_k \left[ \la n_{k\ua}\ra + \la n_{k\da} \ra\right] -
 6\sum_k \la n_{k\ua}n_{k\da}\ra \ee
where the average $\la n_{k\ua}n_{k\da}\ra$ can be well approximated by $\la n_{k\ua}\ra \la n_{k\da}\ra$
since the correlated part of second order processes creating singlets
 at the same $k$ is very small compared to first order contribution.
Therefore the knowledge of $\la n_k\ra$, e.g. from measurement of the excitation density can yield the information necessary
 for the estimation of $\sum_k \la \mathbf{\mathcal{S}}_k^2 \ra$.

Electron spins are decorrelated by random spin flip processes in semiconductors.
It should be advantageous to use samples with
long spin lifetime, such as in  $n$-type bulk GaAs \cite{kikkawa}.  Spin flip times of the photo-excited holes are much faster compared to the
electrons  \cite{fabian-rmp}, and therefore their contribution to the collective spin correlations should be
small.

We would like to thank I. Bar-Joseph and I. Neder for valuable discussions.

\end{document}